\documentclass[conference]{IEEEtran}
\makeatletter
\def\ps@headings{%
\def\@oddhead{\mbox{}\scriptsize\rightmark \hfil \thepage}%
\def\@evenhead{\scriptsize\thepage \hfil \leftmark\mbox{}}%
\def\@oddfoot{}%
\def\@evenfoot{}}
\makeatother
\usepackage{color, soul}
\usepackage{}
\usepackage{tipa}
\usepackage{bbm}
\usepackage{color}
\usepackage{mathrsfs}
\usepackage{amsfonts}
\usepackage{cite,url,subfigure,epsfig,graphicx}
\usepackage{amssymb,amsmath,bm,makecell}
\usepackage{indentfirst}
\usepackage{float}
\usepackage{algorithm}
\usepackage{algpseudocode}
\usepackage{algorithmicx}

\usepackage{tabulary}
\usepackage{booktabs}
\usepackage{textcomp}
\usepackage{multirow}
\usepackage{tabu}
\usepackage{booktabs}
\usepackage{threeparttable}

\IEEEoverridecommandlockouts
\usepackage{times,verbatim,amsfonts,amsmath,color}

\begin{document}

\title{
Metaverse CAN: Embracing Continuous, Active, and Non-intrusive Biometric Authentication
\author{
\IEEEauthorblockN{Hui Zhong{†}, Chenpei Huang{†}, Xinyue Zhang and Miao Pan}
}
\thanks{H. Zhong, C. Huang and M. Pan are with the University of Houston; X. Zhang is with the Kennesaw State University. †These authors contributed equally to this work and share the first authorship.}
}

\maketitle\thispagestyle{empty}\maketitle\pagestyle{empty}

\begin{abstract}
The Metaverse is a virtual world, an immersive experience, a new human-computer interaction, built upon various advanced technologies. How to protect Metaverse personal information and virtual properties is also facing new challenges, such as new attacks and new expectations of user experiences. While traditional methods (e.g., those employed in smartphone authentication) generally pass the basic design considerations, they are repeatedly reported to be either unsafe or inconvenient in the Metaverse. In this paper, we address this discrepancy by introducing CAN: a new design consideration especially for the Metaverse. Specifically, we focus on the legacy and novel biometric authentication systems and evaluate them thoroughly with basic and CAN considerations. We also propose an ear-based method as one example of CAN systems. To conclude, a continuous, active and non-intrusive biometric system is suggested for Metaverse authentication for its capability in continuous sessions, against imposters, and immersive experience.

% Because of the wide range of applications, Metaverse stores a large amount of user information. Experts have proposed various new user authentication methods, while this paper focuses on biometric methods with significant advantages in Metaverse. Existing biometric evaluation criteria are specific to multiple domains and applications which are the basis for authentication evaluation. We narrow the scope of this criterion to specifically address the specific Metaverse environment. In this paper, we propose a new biometric consideration CAN: Continuous, Active and Non-intrusive only for Metaverse. We define these three properties and describe the CAN system detailed. Next, existing biometric methods are classified and evaluated based on basic considerations and CAN considerations. Specifically, we introduced a novel transfer-learning and data augmentation method for suggested ear-based methods. Finally, potential future biometric methods are guided. In summary, we propose a novel set of CAN considerations in the hope of better evaluating biometric methods in Metaverse.

\end{abstract}

\section{Introduction \label{introduction}}

Metaverse, built upon modern computer science and technologies, is introducing a completely simulated digital world to its users\cite{ning2023survey}. In this world, avatars and virtual objects all allow natural interactions (e.g., gesture and speech), which creates a real and immersive experience. The magic of Metaverse is founded on numerous interdisciplinary technologies, including computer vision, the Internet of Things, wireless communication, and so on. For instance, advanced video rendering on head-mounted devices (HMD) is the key to providing a 360-degree experience, allowing users to see and interact with objects from any angle. The sensing system embedded in HMD or joysticks aggregates multi-modal data. and subsequently, estimates user feedback. Also, the rapid development of wireless communication (e.g., 5G NR and WiFi 6) supports high-speed, low-latency, and energy-efficient Metaverse services. Therefore, thanks to these advanced and constantly growing technologies, we hopefully witness Metaverse reshaping the style of life, including entertainment, eLearning, remote working, telemedicine, and even military applications. At the same time, it also raises concerns about Metaverse authentication to secure the user from property and privacy leakages. However, simply applying current authentication methods to Metaverse is either insecure or cumbersome. We emphasize three unique characteristics of Metaverse authentication: 
\textbf{C1:} The user input for Metaverse is significantly different.
\textbf{C2:} Attacks and countermeasures focus on Metaverse onboard sensors.
\textbf{C3:} Pairing with a smartphone/computer is no longer required in future Metaverse.

% Metaverse is a virtual world closely connected to the real world, allowing users to participate in various activities after wearing specific devices. It is becoming the next stage of internet research. Metaverse integrates various technologies such as augmented reality, virtual reality, the Internet of Things, and machine learning. The Internet of Things is the foundation of the metaverse. It connects intelligent devices and sensors from the real world with the virtual world, enabling information exchange between different environments. VR, AR, and XR technologies enable the metaverse to integrate the real and virtual worlds, truly achieving human-computer interaction. The rapid development of 5G also provides support for the low-power metaverse. The metaverse has brought many changes to people's lives and continuously provides users with immersive experiences. Users can use virtual identities to engage in communication, convey information, and even collaborate to complete an activity. For example, users from around the world can participate in meetings as in real life. However, the large amount of user information stored in metaverse can lead to security issues. User authentication has become a very important aspect.

We elaborate on those new challenges by taking the most common password-based authentication as an example. Although typing passwords with a keyboard or touchscreen can be easier, doing so on AR/VR headsets is not user-friendly. One solution is to use a joystick to enter passwords on a virtual keyboard, which takes more time to authenticate\cite{stephenson2022sok}. When it comes to new attacks, the HMD equipped on the user's head will block the view of the surroundings. Unfortunately, the legitimate user's authentication process could be exposed to an attacker, who will launch shoulder-surfing or imitation attacks via recording. Then, we discuss how the new sensors on Metaverse devices trigger the arms race between attacker and defender. For instance, high-end motion sensors are now embedded in commercial off-the-shelf (COTS) headsets (e.g., Meta Quest) to enable gestures using head movement.
On the other hand, a side-channel attack using motion sensor data can interpret a user's speech and speech-related information (e.g., gender, age, etc.) when the legitimate user utilizes voice commands or online chatting. In this attack, motion sensors are backdoored by the attacker to record concurrent facial vibrations related to human speech \cite{li2023security}. One should also notice that Metaverse devices tend to become stand-alone without auxiliary devices. In other words, authentication aided by mobile phones such as scanning QR codes, will eventually fade out. 

\begin{figure}[t]
\centering
\includegraphics[width=0.48\textwidth]{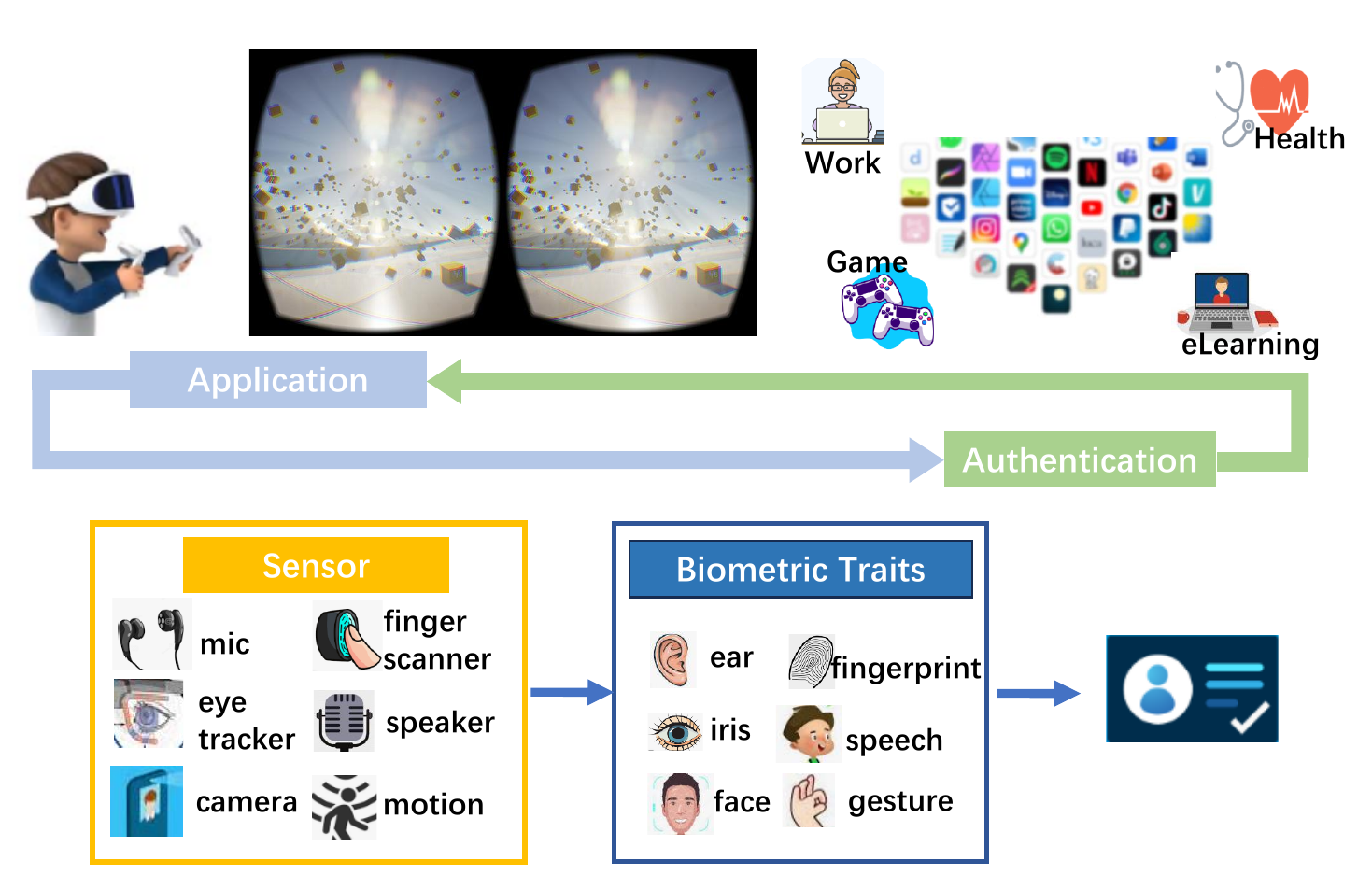}
\caption{Metaverse application and authentication. When an application makes an authentication request, the Metaverse device utilizes onboard sensors to collect the target user's biometric signal for user identification or verification.}
\label{fig:metaverse teaser}
\end{figure}

In light of all the discussions above, it is crucial for the research community to develop and assess potential successful Metaverse authentication methods. According to \cite{stephenson2022sok}, physical and behavioral biometrics are likely to be widely accepted in the era of the Metaverse, shown in Fig.~\ref{fig:metaverse teaser}. The basic biometric authentication designs consider good accessibility, security, and usability. However, it has been repeatedly reported that \textit{not all} biometrics that perform well with these evaluations are suggested for Metaverse authentication. Therefore, the following question is \textit{which biometrics will succeed in Metaverse and how?} Along with the basic design considerations, we introduce CAN: \textbf{C}ontinuous, \textbf{A}ctive, and \textbf{N}on-intrusive authentication as new criteria for Metaverse. The continuous authentication authorizes the entire Metaverse session and minimizes the window of vulnerability. Active authentication aims to detect attackers who utilize spoofing samples or signal injection to bypass the recognition module. Finally, the non-intrusive design will release the requested user effort when adopting continuous and active authentication.

% In \cite{stephenson2022sok}, many promising biometrics methods are discussed, which are all considered to have good accessibility, security, and usability. Thus, we narrow it down to discuss ``biometric for Metaverse'' and endeavor to answer the following question: \textit{which biometrics will succeed in Metaverse and how.} The scope of this paper includes general (e.g., smartphone-employed) methods: fingerprints, face, speech, and so on, Metaverse-friendly methods: iris, gait, etc., and emerging eye-related methods, ear-related methods, and interaction methods. In this paper, we first review the basic criteria of a usable biometric system. Next, \textbf{C}ontinuous, \textbf{A}ctive, and \textbf{N}on-intrusive criteria are clarified and proposed as a new design consideration for Metaverse authentication. Finally, we discuss the CAN system implementation and propose a suggested ear-based Metaverse authentication system.

The paper structure is as follows. In Section~\ref{sec:background}, we introduce the background of Metaverse authentication. Then, we discuss CAN in Section~\ref{sec: CAN description}. In Section~\ref{sec: CAN system}, we present the implementation rules, evaluations, and our design of a recommended CAN system as an example. Finally, we conclude this paper in Section~\ref{sec:conclusion}.

\section{Background \label{sec:background}}

\subsection*{Biometric Methods in Metaverse}

Metaverse can be defined from many perspectives. Technically, it is a computer science innovation to provide spatial and immersive experiences, based on Virtual Reality, Augmented Reality, and Extended (or Cross) Reality (AR/VR/XR)\cite{ning2023survey}. Besides, economic, social, and identity systems, such as blockchain technology, are also involved. Three fundamental characteristics have been discussed in Metaverse survey\cite{ning2023survey}, namely multi-technology, sociality and hyper spatiotemporality. Finally, from the user's perspective, the key to Metaverse is to enable fully immersive human-computer interaction. Therefore, our focus on the Metaverse-friendly biometric authentication should leverage the emerging technology and follow the original intention of Metaverse. In our scope, we treat the Metaverse platform as a multi-modality sensing platform, which is capable of collecting human physiological and behavioral data for user-level pattern recognition.

Biometric authentication has become one of the most prevalent authentication adopted in computer systems, smartphones, and smart home Internet of Things. When operating biometric authentication, an individual should provide a sample (e.g., fingerprint/face image) and the system must compare this sample with a registered template \cite{li2023security}. A standard biometric system includes data acquisition, a feature extractor, a template generator, and a matching engine. It requires two sub-processes to authenticate a user: the enrollment process and the authentication process. In general, a successful biometric authentication should possess user-discriminative biometric identifiers, which can identify a unique user and be resilient to replay and spoofing attacks.

In recent studies, tremendous efforts have been dedicated to the marriage between Metaverse and biometrics. The reasons are three-fold. First, the Metaverse platform already shows its increasing sensing capacity based on novel embedded sensors. Second, Metaverse devices are designed to process multi-modal user interaction, which can be made into behavioral biometric methods. For instance, Gaitlock proposed by Shen et al. \cite{shen2018gaitlock} utilized user gait characteristics to verify a registered user. Third, for the stand-alone HMD platforms, since the devices are often used for a long period of time, vital signs and liveness-related signals can be easily picked up and utilized for continuous authentication. Skullconduct \cite{schneegass2016skullconduct} harnessed the bone-conduction earphones and microphones on Google Glass and recognized the legitimate user based on the unique reflective response. In a nutshell, while it requires extra development efforts, using biometrics can cause extremely low user participation and distraction, which is emphasized in the Metaverse.

% The movement of the user's eyes or head is minute, while most of behavioral biometric methods require the user to actively make substantial movements. Shen et al. proposed a system to discriminate users based on their gait characteristics. This system does not require special equipment, but it is difficult to implement in public and is vulnerable to imitation attacks. The physical biometric method is based on the user's biometric characteristics, of which face and fingerprint recognition methods are widely used in daily life. Earprint is based on the reflected signal of the sound after it has passed through the user's body. Skullconduct is based on the reflected signal of the sound after it passes through the user's skull. These systems are active and requires very little effort from the user.

\subsection*{Basic Biometric System Consideration}
% In this subsection, the following considerations are useful to assess one biometric authentication method \cite{stephenson2022sok}. Conventionally, the \textit{Accuracy}, \textit{Security}, \textit{Deployability}, \textit{Accessibility}, and \textit{Convenience} are extensively discussed in traditional systems.

\noindent \textbf{Accuracy.} Accuracy measures the probability that a biometric authentication system succeeds in authorizing the legitimate user and rejecting a stranger or attacker. The commonly employed metrics are balanced accuracy, equal error rate (EER), and Reactive Operating Characteristics (ROC) as a visual evaluation.
% Specifically, in the authentication process, True Positive Rate (TPR) and False Negative Rate (FNR) characters the probability that a genuine input pattern is accepted/rejected after matching with the pattern in the database; True Negative Rate (TNR) and False Positive Rate (FPR) characters the probability that a non-matching input is rejected/accepted by the matching engine. Furthermore, Equal Error Rate (EER) is the rate when the FNR equals to the FPR, where a lower value indicates better general performance. Reactive Operating Characteristic (ROC) is another metric to show the trade-off between two kinds of errors. By plotting the curve, a bigger Area under the ROC Curve (AUC) indicates a more accurate result.

\smallskip
\noindent \textbf{Security.} 
Security measures the system's resilience against several vulnerabilities. In the context of biometric authentication, target attacks such as spoofing attacks are emphasized. For example, face recognition systems can be easily bypassed by providing a face photo from the victim user. We will address the threat models in the next subsection.

% Although biometrics is unique and easy to use, various attacks pose threats to using biometrics for user authentication. For example, the spoofing attack can be easily launched when the legitimate user's biometric identifier is exposed to an attacker. To name a few, face recognition systems can be easily bypassed by providing a face photo from the victim user. The speaker recognition system is vulnerable to replay attacks or more advanced voice conversion and speech synthesis attacks. For behavioral biometrics (e.g., gait), an attacker can learn to imitate the user's authentication process by observation and recordings. Nowadays, when deep learning algorithms are involved in biometric authentication, the new attack dimension from deep learning vulnerability is also under active discussion.
 
\smallskip
\noindent\textbf{Deployability.} It is noteworthy that \emph{not all} biometric systems can be readily implemented on existing platforms. The difficulties originate from the lack of software/hardware development and supporting tools, which presents significant challenges to the platform developer. 
% Typically, while the fingerprint scanner is on the rise in usage, fingerprint biometrics can be defined as deployable. On the other hand, event-related potential (ERP) brain waves, which requires brain-computer interfaces, are less deployable.

% ployable, which means the approach cannot follow a proven framework or platform. Some of the newer methods also require the user to wear additional sensors, such as ERP brain waves that require the use of a specific brain-computer interface.

\smallskip
\noindent\textbf{Accessibility.} The accessibility is explained as how the biometric system can be readily operated by users with disabilities, such as visual, hearing, speech, mobility, and cognitive. In other words, a biometric method loses certain accessibility if certain user interactions are complicated for individuals with disabilities.

% \smallskip
% \noindent\textbf{Convenience.}
% Convenience is reflected in how much effort the user needs to make. Common biometric methods are passive which requires user-initiated behavior, such as face recognition, fingerprint recognition, and behavioral biometrics. The new physical biometrics can basically do the system proactively. In other words, the user doesn't need to perform laborious actions.

%The usability of biometric methods can be divided into deployability, accessibility, and convenience.
%Convenience is reflected in how much effort the user needs to make. Common biometric methods are passive which requires user-initiated behavior, such as face recognition, fingerprint recognition, and behavioral biometrics. The new physical biometrics can basically do the system proactively. In other words, the user doesn't need to perform laborious actions. Accessibility is reflected in whether biometric methods are disabled-friendly. Some biometric methods can restrict access to specific groups of people, for instance, speech-based recognition is not friendly enough for users with language impairments. Finally, it is worth noticing that biometrics is usually not well-deployable, which means the approach cannot follow a proven framework or platform. Some of the newer methods also require the user to wear additional sensors, such as ERP brain waves that require the use of a specific brain-computer interface.

\subsection*{Threat Model}
% We consider two general attack types in the context of Metaverse authentication: user-present attacks and user-absent attacks. In the former attacks, the attacker's goal is to directly unlock the device/log into an account or gain useful information to do so without notifying the user, who is currently using the Metaverse device. In the latter attacks, the attacker is assumed to have physical access to the target Metaverse device. Although any interactions could be possible, tampering with hardware or physically compromising the authentication system (e.g., sensor/CPU/memory) are out of the scope of this paper.

We identify the following threat vectors which may originally target any computer systems but retain the potential to overcome the Metaverse authentication without the legitimate user's permission. They are, (1) Insider attacks, mentioned in \cite{zhang2018continuous}, this attacks take over the target account and conduct unauthorized actions when the device is left accessible to that insider; (2) Shoulder-surfing attacks, this attack let the attacker observe and illegally record the legitimate user's login effort, which raises serious concerns due to Metaverse employing HMD; (3) Side-channel, attacks leverage physical side-channel to steal or bypass the login effort; (4) Spoofing attacks, attacker generate fake biometric samples to fool the authentication system.

To make it clear, say Alice wants to launch attacks to steal Metaverse user Bob's account. If Alice is an insider (i.e., a co-worker near Bob) and when Bob leaves his Metaverse device on the table, she can make purchases using Bob's account before the login expires. Alice can also record when Bob logs into his device, for example, using voice commands or certain gestures. Additionally, Alice could trick Bob into installing malicious software or grant permissions so that the device can be affected via side channels. The last example is that Alice can generate Bob's facial images or voice samples by replaying them in front of the camera/microphone if Bob decides to use such biometrics for authentication.

\section{CAN: New Consideration for Metaverse \label{sec: CAN description}}
% Why is traditional metric not sufficient/adaptive to new tasks? Why CAN? What is CAN?
We now move forward to discuss the new considerations in the context of biometrics for Metaverse. In a nutshell, developing biometric systems for Metaverse should address the new using habit, immersive experience, and novel user interaction. Therefore, we nominate three considerations: Continuous, Active, and Non-intrusive criteria for Metaverse biometrics. In previous literature, these terms have lacked clear clarification and have sometimes been used interchangeably. The purpose here is to propose suggested definitions for Metaverse authentication systems, shown in Fig.~\ref{fig:system_block}.

\vspace{-1mm}
\subsection*{Continuous vs. One-time Authentication}
%Definition. 
%Why continuous? Suitable for Metaverse Typical example.
Current authentication methods run a one-time authentication and then give full access to the user who passed the login stage. We found it insufficient for long and continuous sessions as the previous study demonstrated. Unfortunately, Metaverse sessions are considered long and continuously receive user input, which leads to the adoption of continuous authentication. Continuous authentication authenticates the user's presence during the session. The biometric traits are continuously or repeatedly captured when triggered by certain events. The captured biometric traits should be continuous physical input (e.g., speech/video) or sampled over session time (e.g., image). The rule of thumb for discrete-time sampled biometric traits is that the acquisition duration and frequency should be comparable to that of a user's operation during the session. Therefore, continuous authentication is deemed as a fusion in time \cite{niinuma2010soft}.

% mostly verify the user's identity at the login stage. Next, the service provider will provide access to the authorized user until the current session ends. Unfortunately, this one-time authentication may be insufficient in the case of Metaverse authentication. As a Metaverse session is longer than others, the attack can be launched \textit{anytime} and have an increased chance of success (since it has more time to prepare). In addition, unauthorized use of Metaverse may lead to severe property and information loss, which also makes it inappropriate to grant the "post-authorized user" full access. Taking the increased vulnerability and cost of failure into consideration, continuous authentication, which authenticates the user's presence continuously during the session \cite{niinuma2010soft}, serves as a key in the Metaverse authentication.

We then answer why continuous authentication is recommended for Metaverse. First, the Metaverse applications mostly require long-time user interaction, which makes them vulnerable to intruders when the post-authorized user is absent. Second, the spirit of continuous and multi-modal human-computer interaction in Metaverse coincides with continuous authentication. Third, as mentioned in \cite{niinuma2010soft}, maintaining a user's authentication state can rely on multi-modal and soft biometrics, which significantly ease the user's effort or concern. This relaxed version of continuous authentication will inspire the Metaverse developer in software/hardware designs.

\vspace{-1mm}
\subsection*{Active vs. Passive Authentication}
%What is Active (challenge-response)
%Benefits: Secure

%Passive definition/clarification
%Limitation.
Active authentication is behind the idea that the system actively asks ``who you are'' instead of passively scoring a user's submission, according to the well-known DARPA proposal \cite{DARPAActiveAuth}. The formal definition of active authentication states ``it is an automated recognition mechanism, which employs active sensors to measure the biometric information for decision-making.'' Adopting active authentication aims to shield the system from the adversary which is possible to bypass the passive authentication system by submitting suspicious samples. The defense strategy in active authentication is to initiate the data acquisition without notifying the user (attacker) so that a false response or non-negligible delay could detect the attacker's presence. 

% Formally, active authentication is an automated recognition mechanism, which employs active sensors to measure the biometric information for decision-making. The goal of active authentication is to prevent unauthorized access from imposters, who are a kind of attackers aiming to misinform the passive authentication system via well-crafted analog signals, e.g., voice spoofing attacks misleading speech verification systems. Dynamic biometrics and challenge-response-integrated (CRI) biometrics are often used to build active authentication systems on mobile devices, and potentially on Metaverse devices. The design details will be discussed in the next section.

The highlighted vulnerability to spoofing attacks and adversary signal injection necessitates active authentication. First, while wearing HMD and operating joysticks, users cannot be fully aware of their surroundings and detect suspicious signal recordings or injections. Furthermore, common biometrics such as face and voiceprint are publicly available for an attacker to launch replay or even more powerful attacks. Therefore, while suitable biometrics are limited for legacy systems such as smartphones, Metaverse's rich sensing capacity broadens the selection pool without worrying about deployability. In fact, the Metaverse active authentication can be realized based on dynamic biometrics and challenge-response-integrated (CRI) biometrics, which are both (partially) supported by current COTS AR/VR devices. The detailed implementation can be found in Section.~\ref{sec: CAN system}.

% can be easily accessed through public social media, which makes o launch attacks with a remarkably high success rate thanks to advanced generative examples or malicious signal injection. Therefore, active authentication embodies significant advantages over passive authentication. In detail, both dynamic biometrics and CRI biometrics manage to detect the intruder based on the fact that the selected biometrics from  the legitimate user can be verified at any time, while the spoofed or injected ones from the imposter require careful preparations. 
\vspace{-1mm}
\subsection*{Non-intrusive vs. Intrusive Authentication}
% Effortless: physically simple to accomplish
% Casual: casual and stealthy to operate in public
% Immersive: authentication while using
We define the last consideration as non-intrusive authentication. Whilst it is also called convenient, unobtrusive, quiet authentication in other literature, we hereby utilize ``non-intrusive'' for a broader meaning. To explore the non-intrusiveness, we conclude a few characteristics based on whether the operation is \textbf{A}ccessory-free, \textbf{E}ffortless, \textbf{C}asual, and \textbf{U}ninterrupted. 

Next, we elaborate on the proposed non-intrusive characteristics in detail. \textbf{A:} The operation is accessory-free if no extra devices are carried or paired in order for authentication. \textbf{E:} The operation is effortless if it is physically simple to finish for most Metaverse users. \textbf{C:} The operation is casual if it is subtle and quiet, or carefree to do in public, without disturbing the surroundings or notifying the attacker. \textbf{U:} The operation is uninterrupted if the authentication is done without breaking into the immersive Metaverse services (except notifications based on the user's personal settings). Notably, although those characteristics focus on different aspects of the easy-to-use concept, all of them rely on careful biometric selection.

In the sense of Metaverse's user experience, non-intrusive authentication by definition is competent for the most user-friendly authentication solution in multiple scenarios, by minimizing user efforts and concerns. From the security point of view, it also shields the legitimate user from attackers by hiding the real authentication attempts. In this way, the attacker will neither be able to collect history login information in the early stage nor to find the correct attacking window even with well-crafted attacks.

\subsection*{Interconnections}
\begin{itemize}
    \item Continuous authentication requires frequency and duration to match with those of user operations.
    \item To minimize the increased burden caused by continuous authentication, active authentication selects biometrics constantly-ready, which can be submitted at any instance.
    \item Non-intrusive authentication retains an immersive experience without distracting the user from current applications.
    \item The continuous, active, and non-intrusive factors are closely related. Also, they can be jointly achieved for a better experience in Metaverse.
\end{itemize}

% In general, 'Continuous' maintains continuous authentication during prolonged use to ensure information security; 'Active' refers to the system's active measurement of biometric information to prevent spoofing attacks; 'Non-intrusive' refers to the degree of accessory-free, effortless, casual, and uninterrupted operation to assess user experience and protect legitimate users. Figure \ref{CAN system.jpg} shows the process of continuous authentication, the composition of proactivity and non-intrusiveness. CAN co-evaluates the biometric methods in the Metaverse, addressing the above question: which biometrics will succeed in Metaverse and how.

\begin{figure}[t]
\centering
\includegraphics[width=0.49\textwidth]{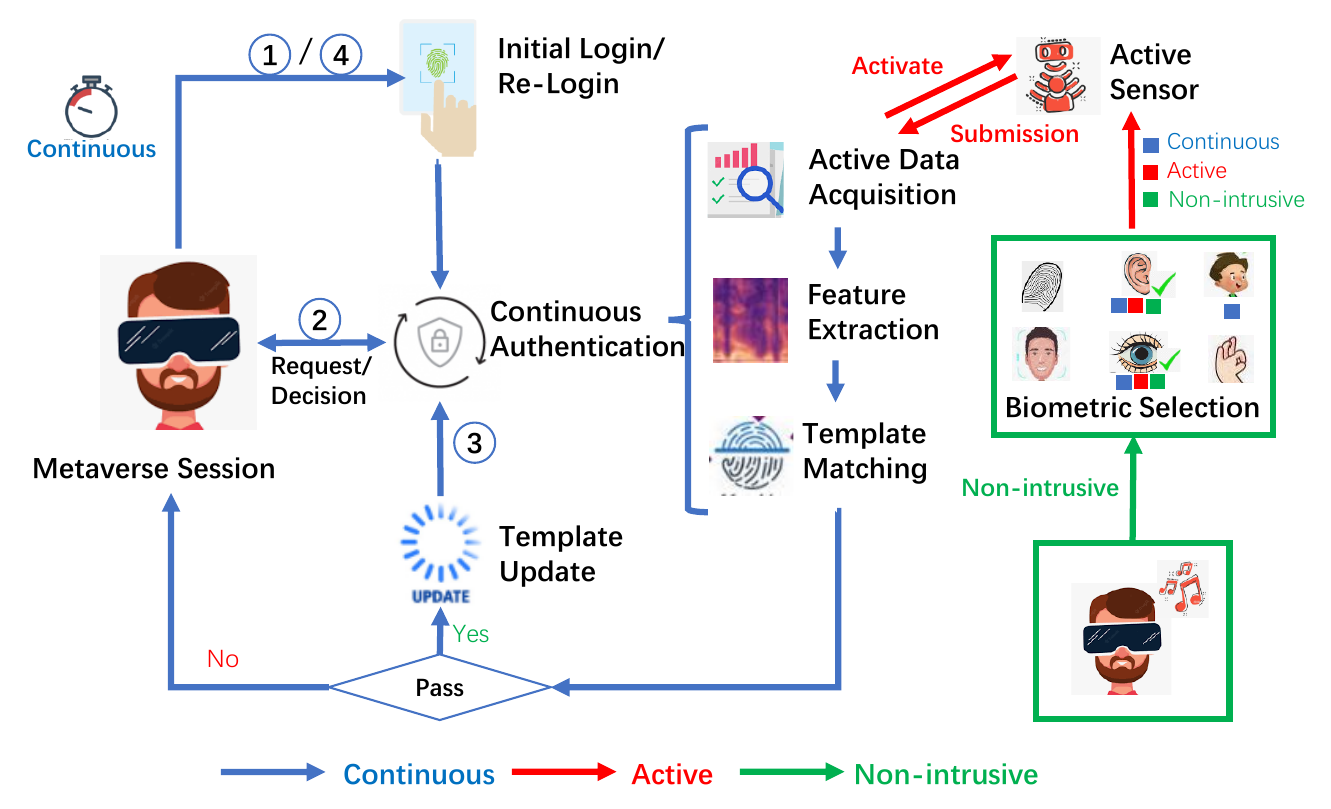}
\caption{Continuous, Active and Non-intrusive system design. (1) Initial Login, (2) Continuous Authentication, (3) Continuous Template Update, and (4) Re-login.}
\label{fig:system_block}
\vspace{-7mm}
\end{figure}

\section{CAN Authentication System \label{sec: CAN system}}
\subsection*{Design and Implementation}
We illustrate the CAN system implementation in Fig.~\ref{fig:system_block}, while details are elaborated as follows.

\smallskip
\noindent\textbf{Continuous.} One natural continuous authentication realization is to incorporate biometrics in cycles. We follow the continuous authentication framework in Koichiro's work \cite{niinuma2010soft}: (1) Initial Login Authentication, (2) Continuous Authentication, (3) Continuous Template Update, and (4) Re-login Authentication. Step 1 uses strict one-time authentication to identify the first presence of the legitimate user. Steps 2 and 3 maintain and update the user's presence status and the corresponding template as the session continues running. When the user fails to prove his/her identity in the middle of the session, step 4 is called for another strict one-time re-login. Notably, the strictness and effort to maintain authenticity can be flexibly designed for deployability and accessibility, since multiple modalities and observations are integrated into the complete verification process. 

\smallskip
\noindent\textbf{Active.}  Dynamic biometrics comprises successively captured physiological biometrics, such as the mobile face, and behavioral biometrics, such as gait, gesture, and gaze. The CRI biometrics, on the other hand, is captured by the system sending stimulus and analyzing the response from the user. Both realizations require active sensors, which are able to acquire signals under the system's control, while the CRI biometrics requires extra actuators to emit stimulus. Unlike the genuine user, the attacker's response to active authentication is likely to be incorrect or significantly delayed. Furthermore, especially for CRI biometrics, a confusing stimulus can be emitted for better attack detection.

\smallskip
\noindent\textbf{Non-intrusive.} Designing an accessory-free authentication system needs select biometrics that can be acquired by the COTS Metaverse devices. To achieve an effortless authentication experience, the data acquisition shall not cost much user cooperation and the system should be resilient to the user's relaxed-and-passable input. When a user wants to authenticate himself/herself casually, the raw signal to be collected (i) is insulated within the user's personal area, (ii) makes unnoticeable (inaudible/invisible) information leakages. Finally, especially considering the Metaverse immersive visual and audible experiences, the selected biometrics require no watching/listening tasks to certain stimuli, which requires the system to authenticate based on the data generated during normal usage.

\subsection*{Current Systems}
% In this section, we will make a classification and evaluation of the current systems. We classified the existing biometric authentication into four categories: smartphone-employed methods, eye-based methods, ear-based methods and interactive methods. We list typical examples (Table.~\ref{table*}) and evaluate them based on basic considerations and novel considerations CAN. The following are detailed descriptions.

% \smallskip
\noindent\textbf{Smartphone-employed methods:} Common traditional biometric methods are fingerprint, facial, gesture, speech and other methods. They are all widely used in current computer systems and have good accuracy rates. However, each of these methods has its own drawbacks in the Metaverse. On the one hand, some methods are less deployable and less accessible. Specific sensors cannot be easily integrated on HMDs, such as external cameras and fingerprint sensors. Gesture recognition is not friendly for people with behavioral difficulties. On the other hand, traditional methods cannot meet almost any of the conditions of CAN. To satisfy the authentication continuity, speech-based recognition requires the user to intermittently initiate the authentication voice, which is impractical in reality and greatly affects the normal immersive experience in the Metaverse.

\noindent\textbf{Eye-based methods:} Eye-based methods refer to sending visually relevant stimuli to trigger a user response or using the uniqueness of the user's iris or periocular. This type of approach is highly accurate and has good deployability with eye-related sensors (e.g. eye trackers, eye-facing cameras). Eye-based methods allow continuous authentication of the user's usage process to be done behind the device and are always active systems. For example, in the design of Eye Movements \cite{zhang2018continuous}, the triggered fixation frequency was set to 0.2 Hz and achieved 93.1\% accuracy, while the user study showed 70\%/62\% of the participants felt this authentication continuously secure and non-intrusive. Another example of ERP Brainwave \cite{lin2018brain} took 4.8 seconds as a one-set password and achieved 95.46\% accuracy.

\begin{table*}[t]
\caption{Evaluation of smartphone-employed and Meteverse-oriented based on basic and CAN consideration}
\vspace{-3mm}
\label{table*}
\resizebox{\textwidth}{!}{
\begin{tabular}{llllllll}
\hline
                                & \multicolumn{4}{c}{\textbf{Basic Consideration}}                                                                                             & \multicolumn{3}{c}{\textbf{New Consideration}}                 \\ \cmidrule(r){2-5} \cmidrule(r){6-8}
\textbf{Description}                                 & \textbf{Accuracy/Participant} & \textbf{Vulnerablity} & \multicolumn{1}{c}{\textbf{Deployability}} & \multicolumn{1}{c}{\textbf{Accessibility}} & \textbf{Continuous} & \textbf{Active} & \textbf{Non-intrusive} \\ \hline
\textbf{Smartphone-employed}                        &                            &                       &                                            &                                            &                     &                 &                        \\ 
password                                            & L                          & shoulder surfing      & screen, joystick                           & action-required                            & N                   & N               & intrusive                      \\
fingerprint                                         & M                          & mimick/synthesize     & fingerprint sensor                         & action-required                            & N                   & N               & EC                     \\
facial                                              & M                          & spoof                 & outer camera                               & NA                                          & N                   & N               & intrusive                      \\
gesture                                             & M                          & mimick                & joystick, IMU, camera                      & action-required                            & N                   & dynamic         & A                      \\
speech                                              & M                          & replay/synthesize     & audio                                      & action-required                            & N                   & dynamic         & intrusive                      \\ \hline
\textbf{Metaverse-oriented}                         &                            &                       &                                            &                                            &                     &                 &                        \\ 
\textit{Eye-based methods}                          &                            &                       &                                            &                                            &                     &                 &                        \\ 
iris image \cite{boutros2020iris}                                         & H: 93.65\%/152             & spoof                 & eye facing cameras                         & vision-required                            & Y                   & dynamic         & AECU                   \\
eye movements
\cite{zhang2018continuous}                                      & H: 93.1\%/30               & mimick                & screen, cameras                            & vision-required                            & Y                   & dynamic         & AC                     \\
ocular image \cite{boutros2020iris}                                       & H: 94.14\%/152             & spoof                 & eye facing cameras                         & vision-required                            & Y                   & dynamic         & AECU                   \\
ERP brainwave  \cite{lin2018brain}                                     & H: 95.46\%/179             & NA                    & screen, wearable BCI                       & vision-required                            & Y                   & CRI             & EC                     \\  \hline
\textit{Ear-based methods}                          &                            &                       &                                            &                                            &                     &                 &                        \\
occlusal sounds \cite{xie2022teethpass}                                    & H: 96.8\%/22               & spoof                 & audio                                      & hearing-required                           & N                  & N               & AC                     \\
ear canal response  \cite{gao2019earecho}                                & H: 94.19\%/20              & mimick                & audio                                      & hearing-required                           & Y                   & CRI             & AECU                   \\
skull vibration  \cite{schneegass2016skullconduct}                                   & H: 97\%/10                 & mimick                & audio, head IMU                            & hearing-required                           & Y                   & CRI             & AEC                    \\
pupil response   \cite{soundlock2023}                                   & H: 99.09\%/32              & NA                    & audio, eye tracker                         & action-required                            & Y                   & CRI             & AEC                    \\  \hline
\textit{Interaction methods} &                            &                       &                                            &                                            &                     &                 &                        \\
gait \cite{shen2018gaitlock}                                               & H: 98\%/20                 & mimick                & head IMU                                   & action-required                            & N                  & N               & A                      \\
throw virtual ball \cite{ajit2019combining}                                  & H: 93.03\%/10              & mimick                & screen, joystick                           & action-required                            & N                  & N               & intrusive                      \\
EMS motion  \cite{chen2021user}                                        & H: 99.78\%/13              & spoof                 & body IMU, EMS                              & action-required                            & N                  & CRI             & intrusive                      \\ \hline
\end{tabular}}
\begin{threeparttable}
\begin{tablenotes}
\item Note L: low accuracy. M: medium accuracy. H: high accuracy. NA: non-applicable. Y: yes. N: no. A: accessory-free. E: effortless. C: casual. U: uninterrupted.
\end{tablenotes}
\vspace{-8mm}
\end{threeparttable}
\end{table*}

\noindent\textbf{Ear-based methods:} Ear-based methods refer to sending a single signal and audio to trigger a response from the user or an authentication signal collected through a microphone. These methods are highly deployable because they require only a microphone and speaker. In other words, ear-based methods do not require additional equipment. Specifically, Earecho \cite{gao2019earecho} only requires the user to passively listen without making an effort, so the user is not disturbed and can be immersed in the activity. EarEcho achieved 98\% accuracy for a 3-second continuous sampling \cite{gao2019earecho}. Soundlock \cite{soundlock2023}, on the other hand, took 7 seconds for a similar accuracy. According to their user survey \cite{soundlock2023}, participants rated the system as non-intrusive at 4.48 out of 5 (5: the most acceptable).

\noindent\textbf{Interaction methods:} Interaction methods refer to the user's active authentication by a specific action or the response of a body part caused by a stimulus. These methods are usually novel behavioral biometric methods, but are difficult to be accepted by users. Interaction methods are not well deployable, because they often require additional equipment (e.g., a joystick for throwing a virtual ball\cite{ajit2019combining}). Interaction methods cannot be used by people with limited mobility. Except for Electricauth\cite{chen2021user}, all other typical methods require the user to initiate authentication and hardly satisfy the AECU (Accessory-free, Effortless, Casual,
and Uninterrupted) of non-intrusiveness at any point. Therefore, it is not applicable to user authentication in the Metaverse.

\subsection*{Our Ear-based CAN Design: An Example}
 Research has demonstrated that human ears exhibit unique patterns that differentiate them from one another. While ear recognition was initially conducted in the visual domain, recent efforts have focused on acoustic sensing methods, which can be carried out by head-wearable devices, e.g. Metaverse headsets. The core of audio-domain ear recognition is to capture the in-ear response: to transmit audio into the ear and receive the reflections bounds by the wall of the ear canal. Existing works have shown these ear-based methods are promising in basic considerations (accuracy, security, deployability, and accessibility), but the performances w.r.t. the new consideration CAN are still unsatisfactory. 

 Since all ear-based acoustic authentication methods need speakers to emit acoustic signals, the choices of the stimuli determine whether the systems are continuous, active, or non-intrusive. Typically, two mainstream acoustic signals for ear canal sensing are designated sensing signals (chirp, white noise, or PN sequence) and regular audio playbacks (podcast, conversation, music, and so on). The designated sensing signals are used for channel sounding, whose auto-correlation is an impulse function. These sensing signals usually cover the entire human auditory frequencies from 20 Hz to 20 kHz. While a good frequency resolution can be attained in this way for significantly accurate performance, the loss of naturalness usually makes listeners uncomfortable, which makes it impossible to build into continuous and non-intrusive authentication systems (although active). On the other hand, utilizing regular audio playbacks is extremely suitable for a continuous and non-intrusive design. However, since it is common to encounter silent segments, activeness is not applied. In addition, estimating ear canal transfer function via band-limited audio playbacks is not as accurate as those using sensing signals.

% \begin{figure}[t] \centering %\hspace{-3em}
%   \subfigure[Flow chart: (1) Chirp sensing for registration. (2) Augmentation and user template creation. (3) Authentication with any displayed audio. (4) Patchwork CAN implementation.\label{fig:ear-based system}]
%   {\includegraphics[width=.46\textwidth]{Fig_metaverse/CAN_system.png}}
  
%   \subfigure[The PSD of CAN audio playback: original (blue) and patched (blue).\label{fig:patchwork}]
%   {\includegraphics[width=.35\textwidth]{Fig_metaverse/CAN_patch.png}}
%   \caption{Proposed ear-based method fulfilling CAN consideration. }
% \end{figure}

\begin{figure}[t]
\centering
\includegraphics[width=0.45\textwidth]{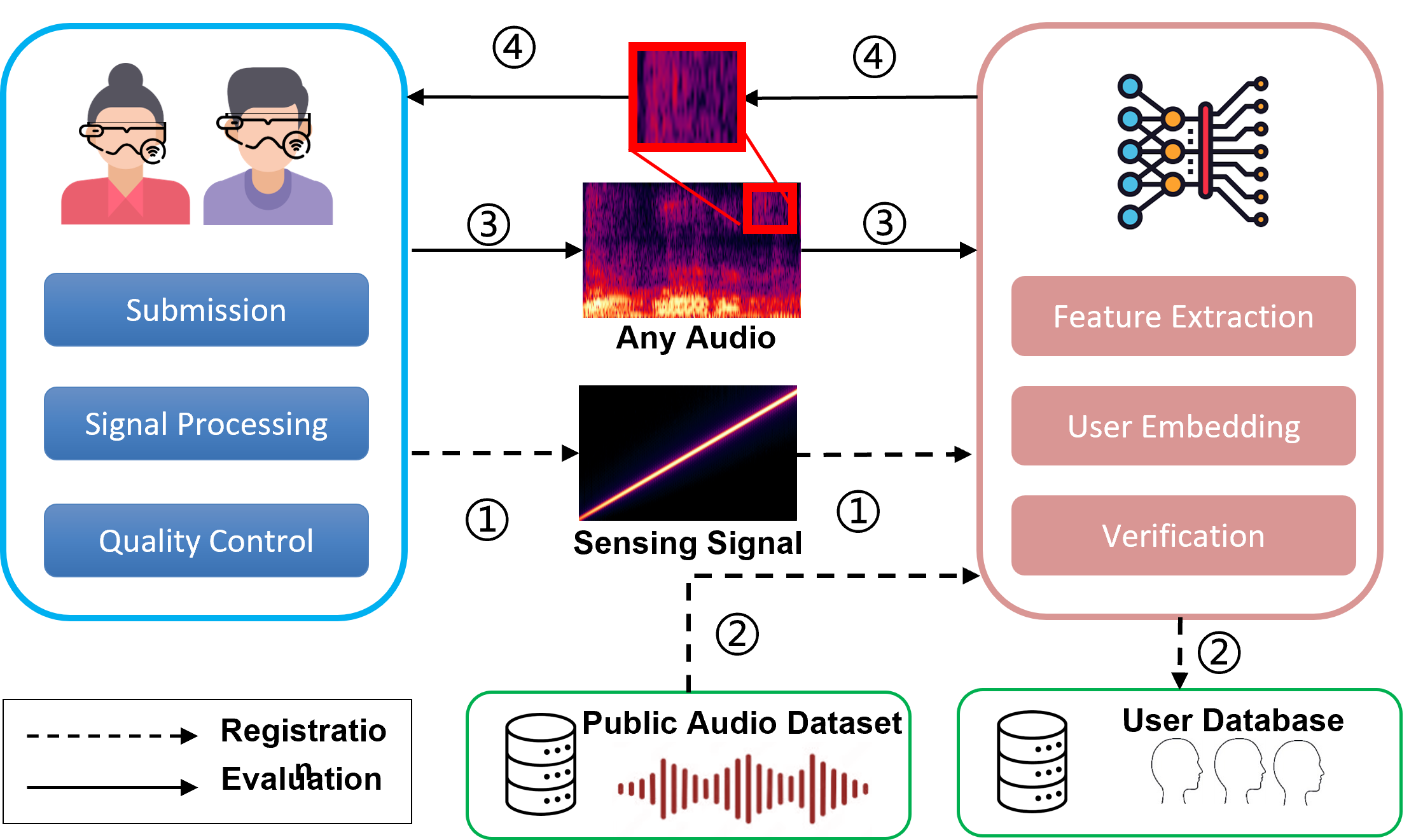}
\caption{Proposed system overview: (1) Chirp sensing for registration. (2) Augmentation and user template creation. (3) Authentication with any displayed audio. (4) Patchwork CAN implementation.\label{fig:ear-based system}}
\vspace{-6mm}
\end{figure}

 We proposed a novel transfer-learning and data augmentation method, that leverages the advantages of sensing- and playback-based ear-based authentication, which can be found in Fig.~\ref{fig:ear-based system}. Notice that in the enrollment (registration) stage, users are assumed to collect the in-ear signals as biometrics in a controlled environment (high quality), and they can tolerate the one-time intrusive experience for a better experience later. Therefore, sensing signals such as exponential chirps are employed for biometric template generation. The next goal is to allow user recognition in the authentication (evaluation) stage via \textit{any} audio playbacks. The key to bridging the gap between the two stages and their selected signals is to use deep learning with data augmentation. Specifically, based on the high-quality ear canal response collected in the enrollment stage, we synthesize a large batch of playback signals by convolving the user's response with public audio datasets. Then, a deep learning model (e.g., ECAPA-TDNN) is trained based on the synthesized data (audio-response pairs) for feature extraction and template generation. In this way, during the evaluation, a similar embedding vector will be generated by feeding the real data (playback-response pairs) to the trained deep-learning model. Therefore, a high matching score is attained if the user is legitimate and vice versa.

 We recommend this method because it surpasses the current ear-based solutions in the CAN aspects. By slightly sacrificing the experience during enrollment, the proposed method is deemed as continuous and non-intrusive with increased accuracy/security. Our next attempt is to achieve activeness, which means the data actively collected at any instance by the system serves as biometric traits. In the context of ear-based authentication, the problem is how to combat \textit{silent segments} and \textit{absent frequencies}. Fortunately, the human auditory system cannot perceive all differences between two similar audio, which can significantly improve the sensing performance without notifying the user. Therefore, we optimize the patchwork to watermark the silent/frequency-missing audio clips and set an ``unnoticeable constraint" based on \textit{minimum audibility curve} and \textit{audio masking effects}. The patchwork is crafted by jointly optimizing the recognition loss function (i.e., AAMSoftmax) and the auditory perception differences (w/ vs. w/o watermark) using gradient methods. To summarize, our proposed ear-based authentication is the first design that is continuous, active, and non-intrusive. We believe similar works can be done by fulfilling the CAN requirement based on various novel biometrics, which is recommended for the Metaverse authentication. 

% \section{Discussion\label{sec:discussion}}

% \noindent \textbf{Cancelable Biometrics.}
% Cancellability is a feature of biometric methods that is gaining attention these days. Cancellability means that when the information is maliciously damaged or stolen by an attacker, using another new incentive, the system can get a new reflected signal to verify the user. Biometric methods with cancellation excel in accuracy, security, continuity, and proactivity, but lack deployability, which affects the user immersive experience. It is temporarily only for specific several biometric methods (e.g., Soundlock\cite{soundlock2023}, BrainPassword\cite{lin2018brain}). BrainPassword\cite{lin2018brain} obtains unique password credentials through event-related potential
% (ERP) brainwave. ERP is an unconscious brainwave response, so it cannot be captured by attackers and has very strong security. However, this new method requires a specialized brain-computer interface and user acceptability has yet to be studied. 

\section{Conclusion \label{sec:conclusion}}
In this paper, we propose a new evaluation consideration CAN for the Metaverse environment. We clarify these three considerations: Continuous, Active and Non-intrusive, presenting suggested definitions to help researchers distinguish between them. We present a detailed description of the CAN system and evaluate the existing smartphone-employed and Metaverse-oriented biometric methods based on the basic and CAN considerations. Finally, a novel transfer-learning and data augmentation method is presented as a suggested ear-based CAN design. We hope that the CAN evaluation consideration could be useful for Metaverse authentication and give some inspiration to researchers.

\section*{Acknowledgement}
The work of H. Zhong, C. Huang, and M. Pan was supported in part by the US National Science Foundation under grants CNS-2107057 and CNS-2318664.

\bibliographystyle{IEEEtran}

\end{document}